\documentclass[amsmath,amssymb,twocolumn,prl,showpacs]{revtex4}
\usepackage[dvips]{graphics,color}
\usepackage{dcolumn}
\usepackage{bm}
\usepackage{graphicx}
\usepackage{epsfig}

\newcommand{\beq}{\begin{equation}}
\newcommand{\eeq}{\end{equation}}
\newcommand{\ba}{\begin{array}}
\newcommand{\ea}{\end{array}}
\newcommand{\beqa}{\begin{eqnarray}}
\newcommand{\eeqa}{\end{eqnarray}}
\newcommand{\bd}[1]{ \mbox{\boldmath $#1$}  }

\newcommand{\APNY}[1]{Ann. Phys. N.Y. {\bf {#1}}}

\newcommand{\NPA}[1]{Nucl. Phys. {\bf A{#1}}}
\newcommand{\PLB}[1]{Phys. Lett. {\bf B{#1}}}

\newcommand{\PRC}[1]{Phys. Rev. C {\bf {#1}}}
\newcommand{\PRL}[1]{Phys. Rev. Lett. {\bf {#1}}}
\newcommand{\PR}[1]{Phys. Rep. {\bf {#1}}}

\newcommand{\RMP}[1]{Rev. Mod. Phys. {\bf {#1}}}

\begin{document}

\title{Hindrance of heavy-ion fusion due to nuclear incompressibility}
\author{\c S. Mi\c sicu\footnote{On leave of absence from  
National Institute for Nuclear Physics, Bucharest, P.O.Box MG6,
Romania}\email{misicu@theor1.theory.nipne.ro} and H. Esbensen}
\affiliation{Physics Division, Argonne National Laboratory, 
Argonne, Illinois 60439, USA}
\date{\today}

\begin{abstract}
We propose a new mechanism to explain the unexpected steep falloff of 
fusion cross sections at energies far below the Coulomb barrier. 
The saturation properties of nuclear matter are causing a hindrance to large 
overlap of the reacting nuclei and consequently a sensitive change of the nuclear potential 
inside the barrier. We report in this letter a good agreement 
with the data of coupled-channels calculation for the  $^{64}$Ni+$^{64}$Ni 
combination using the double-folding potential with M3Y-Reid effective $N-N$ 
forces supplemented with a repulsive core that reproduces the nuclear 
incompressibility for total overlap. 
\end{abstract}
\maketitle

In recent years a new phenomenon observed in the  fusion of several medium-heavy dinuclear 
systems  \cite{jiang02,jiang04a,jiang04b,jiang05a,jiang05b} has presented a challenge to the
theoretical understanding of the reaction mechanism at deep subbarrier energies.
From the inspection of capture cross sections for various combinations
of colliding systems such as $^{58}$Ni+$^{58}$Ni, $^{64}$Ni+ $^{64}$Ni, 
$^{60}$Ni+ $^{89}$Y, $^{64}$Ni+$^{100}$Mo, $^{90}$Zr+$^{90}$Zr, 
$^{90}$Zr+$^{89}$Y and $^{90}$Zr+$^{92}$Zr, 
an unexpected hindrance was observed at sub-barrier bombarding energies
below a certain threshold $E_s$ that varies from system to system.
The authors of Ref. \cite{jiang05b} remarked that 
the exploration of the hindrance phenomenon is only in its initial stage and an
identification of the underlying physical cause is still missing. The standard theoretical 
approach to treat capture reactions, i.e., the quantum tunneling through the relative 
barrier of the 
dinuclear system coupled to different low-lying collective degrees of freedom 
(vibrations and rotations) was unable to explain the steep falloff in the 
cross sections. 
The coupled-channels (c.c.) approach failed to describe the dynamics of the fusion process 
below a certain value of the bombarding energy. 
The source of this phenomenon is not only interesting for the understanding of the reaction 
mechanism but may have essential consequences for nuclear reactions that occur in 
stars \cite{jiang05b}. 
It could imply that the synthesis of heavy elements is hindered below a certain 
energy threshold.

To resolve the enigma several hypotheses have been proposed. An issue debated 
by several authors was the large diffuseness needed to fit high-precision fusion 
data \cite{hagino03,newton04}.
A suggestion that is very close in spirit to the present work was
put forward in \cite{dass03}.  The authors made the observation that
potentials such as the Winther-Aky\"uz \cite{browin91} provide
reliable barriers but the fact that they cannot reproduce the data far below the 
barrier is a signature that the ion-ion potential has another form in the inner
part of the barrier. Using simple arguments (no channel coupling, 
WKB calculation of the transmission coefficient) they pointed out that the 
steep falloff in the tunneling probability is related to the disappearance of the 
classically allowed region below a certain energy. If this is true,
it would mean that we are confronted with the existence of a shallow pocket of
the potential inside the barrier.

Another interesting observation \cite{jiang04a} made in relation to the data was 
that the astrophysical $S$-factor, as defined by Burbidge {\it et al.} \cite{burb57},
develops a maximum for the systems enumerated above, which, again, cannot be 
reproduced by c.c. calculations using the deep Winther-Aky\"uz potential. 
For the systems mentioned earlier the maximum occurs at an energy 20-30 MeV above 
the compound nucleus ground state.

We advocate in this work the idea that in order to analyze the fusion data the 
double-folding potentials may  avoid the systematic failure of other potentials.
To reach this goal we have to make some amendments to the schemes that are usually 
employed; see for example Ref. \cite{gontchar04}.
First of all we need to take into account the saturation of the nuclear matter.
Second, the neutron and proton content of the various dinuclear combinations 
have to be included in the potential, a fact which is often 
overlooked or only indirectly accounted for in the Woods-Saxon parametrization.
We show that by properly addressing these issues light can be shed on the  
hindrance in sub-barrier fusion.

There are several facts pointing toward  the existence of shallow 
pockets in the relative heavy-ion potential used to describe 
various nuclear reactions. 
The first evidence came from the discovery of resonant structures 
in collisions of light nuclei, the best known example being the sharp peaks
in the bombarding energy dependence of the excitation curves  
measured in the  $^{12}$C+$^{12}$C scattering by Bromley {\it et al.} \cite{brom60}.
These resonances would then resemble states in a quasi-molecular potential well.

It became clear in the last two decades that the resonant behavior observed 
not only in $^{12}$C+$^{12}$C but also in $^{12}$C+ $^{16}$O,  $^{12}$C+ $^{13}$C, $^{16}$O+ $^{16}$O, $^{16}$O+ $^{24}$Mg, is not an isolated phenomenon occurring 
only in a few a lighter systems.
It persists even in heavier colliding systems, such as the  $^{24}$Mg+ $^{24}$Mg
\cite{zur84} and $^{28}$Si+ $^{28}$Si \cite{bet81}.
Manifestation of clusterization in relation to quasi-molecular pockets is also 
known for various heavy nuclear systems such as cluster radioactivity \cite{rose84}, 
or cold fission of actinides \cite{sand99}.

To simulate the appearance of shallow pockets several recipes have been 
proposed in the past :
a) A Gaussian added to the conventional Woods-Saxon potential was used in 
Refs. \cite{mbm68} and \cite{mich73} in order to fit the reaction cross sections
observed in $^{12}$C+ $^{12}$C,  $^{12}$C+ $^{16}$O and $^{16}$O+ $^{16}$O;
b) potentials computed within the density functional method \cite{blomal67,scheid68} 
and c) the well-known proximity potential (see Ref. \cite{block77} for the 1977  
and \cite{myswy00} for the 2000 versions).

A repulsive core can be also simulated by folding two nuclear density distributions 
with an effective nucleon-nucleon ($N-N$) interaction.
Double folding potentials, without repulsive cores, introduced by Satchler and Love 
\cite{satlov79}, are accurate only in the tail region of the 
nucleus-nucleus potential where the density distributions are only gently
overlapping and thus the assumption of ''frozen density'' is less questionable. 
However this assumption ignores any readjustment due to the mutual excitations 
of the nuclei or the Pauli exclusion principle for strong overlap. 
Density dependent interactions have been 
used in recent times to simulate the saturation of nuclear matter for 
$\alpha$+nucleus scattering (see Ref. \cite{brasat97} and references therein).
For the target-projectile combinations used in the experiments with the ATLAS facility
at Argonne \cite{jiang02,jiang04a,jiang04b,jiang05a}
these double-folding potentials do not necessarily apply because the resulting 
pockets are still too deep and the barriers too thin below a certain energy. 
For these reasons another approach is considered to incorporate the effect of the
nuclear incompressibility. 

To compute the double-folding potential we use the Fourier technique as expounded 
in a previous publication \cite{misgre02}.
Since we are interested in performing dynamical calculations we consider two heavy ions 
with one-body  deformed densities $\rho_1$ and $\rho_2$, subjected to vibrational 
fluctuations, and center of masses separated by the distance $\bd{R}$.
Then the interaction between these two ions can be evaluated
as the double-folding integral of these densities
\beq
V(\bd{R}) = \int d\bd{r}_1 \int d\bd{r}_2
\rho_1(\bd{r}_1)\rho_2(\bd{r}_2)v(\bd{r}_{12},\rho_1,\rho_2),
\label{dfold}
\eeq
where $\bd{r}_{12}=\bd{R}+\bd{r_2}-\bd{r}_1$. 
In the above formula there is allowance
for a density dependence of $v$ as discussed previously.
The central part of the effective $N-N$ potential $v$ contains 
a direct part that also depends on isospin 
since in all the cases of interest $N\neq Z$,
and an exchange part that takes into account the effect of antisymmetrization under 
exchange of nucleons between the two nuclei.
For the density independent part of the effective nucleon-nucleon force we use 
the Michigan-3-Yukawa(M3Y)-Reid parametrization \cite{brasat97,misgre02,bert77}.

The density independent M3Y interactions are correctly predicting the potential for 
peripheral collisions. However reactions sensitive to the potential at smaller distances 
are not reproduced \cite{brasat97}. 
To cure this deficiency the ion-ion potential should also contain a repulsive core.
For the repulsive part we take a double-folding integral  
as in Eq. (\ref{dfold}) but use densities with smaller diffusivity ($a\leq$ 0.4) 
and the $N-N$ interaction is modeled 
by a zero-range form with strength $V_{\rm core}$,
following the suggestion of Ref. \cite{ueg93}.  
In what follows we refer to the used potential (including the Coulomb part 
that is also calculated via the double-folding procedure) as the M3Y+repulsion. 

To derive the properties of the short-range repulsive core, 
we note that an overlapping region with doubled nucleon
density is formed once the distance $R$ between the nuclei becomes less than 
$R_p+R_t$, where $R_p$ and $R_t$ are the nuclear radii along the collision axis. 
The doubling of the density increases the energy of the
nucleons in the overlapping region.
In the case of complete overlap (for $R=0$) the increase of the interaction
energy per nucleon is, up to quadratic terms in the normal density, 
\beq
\frac{\Delta V}{A_p}=\rho_0^2\left.
\left (\frac{\partial^2 B(\rho)}{\partial\rho^2}\right )\right|_{\rho=\rho_0}
=\frac{K}{9},
\eeq
where in the last equality we use the proportionality of the incompressibility 
$K$ of normal nuclear matter to the curvature of the energy per nucleon \cite{eisgre87}. 
Since $K$ is usually not measured directly but deduced from isoscalar giant monopole 
or dipole resonances, and since there are conflicting 
results coming from the cross sections of the corresponding experimental data
(see \cite{shlo01} and references therein) we use instead of a universal value the predictions of the Thomas-Fermi model \cite{myswy98} as a function of the relative neutron excess $\delta=(\rho_n-\rho_p)/\rho$ of the compound (fused) nucleus. 
Eventually the strength of the repulsive core, $V_{\rm core}$, is determined by
assuming that $\Delta V$ must be identified with the value of the heavy-ion
interaction potential at the coordinate origin $R=0$.

The parameters of the Fermi-Dirac density distribution entering in the double-folding  
potential (\ref{dfold}) are calculated as follows:
For the proton density distribution we use the parameters resulting from a
compilation of elastic electron scattering  data \cite{vries87}, 
i.e. $r_{0p}$=1.065 fm
and $a_p$=0.575 fm. For the neutron distribution we took the liberty to vary the 
parameters
under reasonable limits, taking into account the following constraints:
a) we deal with a moderately neutron rich nucleus and b) the barrier of the 
M3Y+repulsion potential should be close to the one predicted by the 
Winther-Aky\"uz potential, since this potential gives a good description of the 
data in the barrier and high-energy region \cite{jiang04b}. Thence we take 
$r_{0n}$=1.085 fm and $a_n$=0.534 fm. To obtain the incompressibility of $K$=228 MeV, 
predicted by the Thomas-Fermi model for the compound nucleus $^{128}$Ba \cite{myswy98}, 
we use a strength of $V_{\rm core}$=496 MeV, and a diffuseness of the repulsive 
part of the density distribution of $a\approx$0.4 fm for both protons and neutrons.

\begin{figure}
\centerline{\epsfig{figure=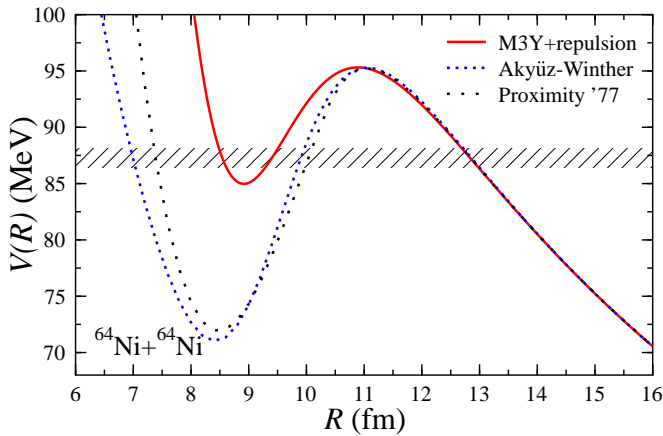,width=0.54\textwidth}}
\caption{Various spherical ion-ion potentials for $^{64}$Ni+$^{64}$Ni.
The solid curve is the potential employed in the present work.
The curve with small dashes is the Winther-Aky\"uz potential used in 
\cite{jiang02,jiang04a,jiang04b,jiang05a,jiang05b}.
The dashed strip corresponds to experimental 
boundaries of the threshold energy $E_s$.} 
\label{pot_ni64_ni64}
\end{figure}

To illustrate the shape of this potential for a case of interest we compare in 
Fig. \ref{pot_ni64_ni64} the spherical heavy-ion potential for the symmetric 
dinuclear system $^{64}$Ni+$^{64}$Ni to different potentials that have been used in 
the past to explain fusion data (Winther-Aky\"uz \cite{browin91}) or fusion 
barriers (Proximity '77 \cite{block77}).
The thickness of the dashed region in Fig. \ref{pot_ni64_ni64} reflects the
uncertainty in the energy $E_s$, where the experimental $S$-factor has a maximum.

As a study case we consider the system $^{64}$Ni+$^{64}$Ni and 
treat the fusion dynamics within the c.c. method 
with linear and quadratic couplings in the quadrupole and octupole 
vibrational amplitudes \cite{esb04}. The c.c. equations,
which include couplings to the low-lying $2^+$ and $3^-$ states,
mutual excitations, and two-phonon quadrupole excitations,
are solved in the rotating frame approximation with the usual outgoing-wave
boundary conditions at large distances. Ingoing-wave boundary conditions
are imposed at the radial separation where the heavy-ion potential develops
a local minimum, $V_{\rm min}\approx 85$ MeV. 
As a result of this scheme, the calculated cross section will vanish
for center-of-mass energies $E\leq V_{\rm min}$. Naturally one would 
expect a non-zero cross section even below $V_{\rm min}$, since the 
compound nucleus can in principle be formed down to a threshold
energy of $E\approx 49$ MeV.

\begin{figure}[t]
\center{\epsfig{figure=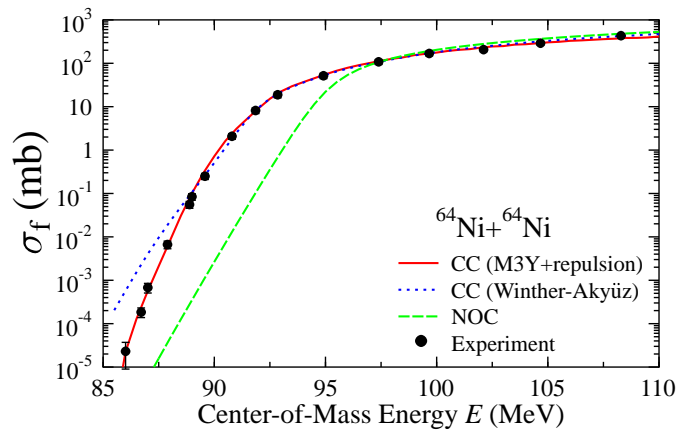,width=0.52\textwidth}}
\caption{Experimental fusion excitation function for the system 
$^{64}$Ni+$^{64}$Ni \cite{jiang04b} is compared to various 
calculations described in the text.}
\label{sig_ni64_ni64}
\end{figure}

In Fig. \ref{sig_ni64_ni64} the experimental excitation function of the fusion 
reaction $^{64}$Ni + $^{64}$Ni $\rightarrow ^{128}$Ba is compared with the results 
using the Winther-Aky\"uz potential as in Ref. \cite{jiang04b} (dotted line) and 
with the M3Y+repulsion potential (solid line), in both cases using the same 
recipe for the c.c. calculations. The cross section for no couplings
is also shown (dashed curve) when the Winther-Aky\"uz potential is used. 
It is seen in Fig. \ref{sig_ni64_ni64} that the agreement with the data, when 
using the M3Y+repulsion potential, is 
much better than the one provided by the Winther-Aky\"uz potential.
The ''M3Y+repulsion'' excitation function has the right shape, not only 
because the potential attains a higher-lying minimum but also because 
the curvature of the barrier is different, 
producing a thicker classically forbidden region.   
Taking all the experimental data points into account we obtain a minimum 
$\chi^2$ per point of $\chi^2/N=10.1$ for an overall energy shift $\Delta E$=0.9 MeV 
of the calculated cross sections when the Winther-Aky\"uz potential is used. 
For the M3Y+repulsion we get a much better result with a minimum  
$\chi^2/N=0.86$ for an energy shift of  $\Delta E$=0.16 MeV.

\begin{figure}[t]
\center{\epsfig{figure=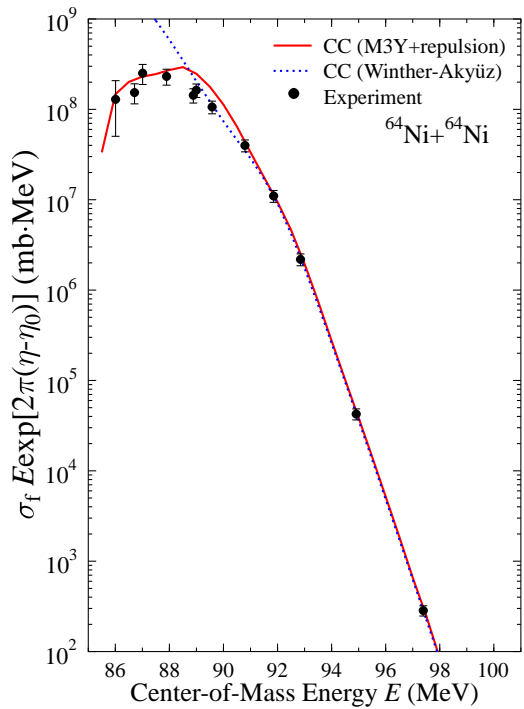,width=0.41\textwidth}}
\caption{The experimental $S$-factor of $^{64}$Ni+$^{64}$Ni \cite{jiang04b} 
(solid points) is compared to the coupled-channels calculations performed with 
the M3Y+repulsion (solid curve) and the Winther-Aky\"uz (dashed curve) potential.}
\label{sfa_ni64_ni64}
\end{figure}

The $S$-factor representation of the $^{64}$Ni+$^{64}$Ni data (solid points) is 
compared in Fig. \ref{sfa_ni64_ni64} with the two c.c. calculations,
based on the  M3Y+repulsion potential (solid curve) and the Winther-Aky\"uz potential
(dashed curve).
The clear maximum in the $S$-factor is reproduced only by the  M3Y+repulsion
potential.
At this point one should recall the experience gained in the sixties on molecular 
resonances.    
As shown in \cite{patt71} the $S$-factor exhibits a sequence of quasi-molecular resonances
for lighter systems. 
In the present case we get a maximum that is too broad to be assigned to a resonance, the 
curvature in the $S$-factor being explained by the shallow pocket in the potential.    
The maximum of the theoretical curve corresponds approximately to the maximum of the 
data points.

A similar study was performed on another
case, $^{58}$Ni+$^{58}$Ni, for which data are available from an older experiment with
the smallest cross sections in the range of mb \cite{beck82}. Even in this case we confirm 
the trend of the excitation functions, namely a faster decrease than expected due to the 
existence of a repulsive core. 


Thus, the understanding of the experimental data, at least for  
the case $^{64}$Ni+$^{64}$Ni, requires 
a modified shape of the potential inside the barrier. 
Both a thicker barrier and a shallower pocket are needed to explain the measured 
cross sections and these features are naturally explained by the M3Y+repulsion
heavy-ion potential. 
If the actual potential pocket would be deeper than the potential used by us, e.g.
the double-folding with density dependent M3Y interactions \cite{brasat97}, then one 
would expect within the energy range between the barrier
top and the threshold energy $E_s$ the appearance of a resonant structure of the form 
encountered for lighter quasi-molecular systems \cite{patt71}. 
Apparently there is no indication of resonant structures in the excitation functions
of $^{64}$Ni+$^{64}$Ni or in the other systems studied in 
Refs. \cite{jiang02,jiang04a,jiang04b,jiang05a,jiang05b}.

One of the authors (\c S.M.) is grateful to the Fulbright commission 
for support through a research scholar fellowship and the hospitality of 
the Physics Division at Argonne National Laboratory where this work was completed. 
H.E. acknowledges the support of the U.S. Department of Energy, Office of Nuclear Physics, 
under contract No.W-31-109-ENG-38. H.E. is grateful to C.L. Jiang, 
B.B. Back, R.V.F. Janssens and K.E. Rehm for a long-term  collaboration that served to clarify
the fusion hindrance phenomenon. Both authors are indebted for their constructive
comments on this article.

\end{document}